\documentstyle[12pt]{article}
\newlength{\myleftmargin}
\newlength{\paperwidth}
\begin{document}
\begin{titlepage}
\begin{center}
{\Large {\bf Semileptonic Decay of $B$-Meson into $D^{**}$}}\\ 
{\Large {\bf and}}\\
{\Large {\bf the Bjorken Sum Rule}}
\end{center}
\vspace{0.3em}
\begin{center}
{\sc Masahisa~Matsuda}~~and {\sc Tsuyoshi~B.~Suzuki}${}^*$
\end{center}
\begin{center}
\sl{Department of Physics and Astronomy, Aichi University of 
Education,}\\
\it{Kariya Aichi 448, Japan}
\end{center}
\begin{center}
${}^*$\sl{Department of Physics, Nagoya University,}\\ \it{Nagoya 
464-01, Japan}
\end{center}

\vskip 4cm
%%%%%%%%%%%%%%%%%%%%%%%%%%%%%%%%%%%%%%%%%%%%%%%%%%%%%%%%%%%%%%%%%%% 
%%%%%%%%%%%%%%%%%%%%% ABSTRACT %%%%%%%%%%%%%%%%%%%%%%%%%%%%%%%%%%%% 
%%%%%%%%%%%%%%%%%%%%%%%%%%%%%%%%%%%%%%%%%%%%%%%%%%%%%%%%%%%%%%%%%%% 

\begin{abstract}

We study the semileptonic branching fraction of $B$-meson into higher 
resonance of charmed meson $D^{**}$ by using the Bjorken sum rule and 
the
heavy
quark effective theory(HQET).
This sum rule and the current experiment of $B$-meson semileptonic 
decay into $D$ and $D^*$ predict that the branching ratio into
$D^{**}l\nu_l$ is about 1.7\%.
This predicted value is larger than the value obtained by various 
models.
\end{abstract}
\end{titlepage}

%%%%%%%%%%%%%%%%%%%%%%%%%%%%%%%%%%%%%%%%%%%%%%%%%%%%%%%%%%%% 

It is well known that the heavy quark effective theory (HQET) is a 
very useful method to study physics of hadrons containing a heavy 
quark\cite{IW}.
For example, the HQET is used to determine Kobayashi-Maskawa matrix 
element $|V_{cb}|$,
>from experiment of semileptonic decay $B\to D^* l\nu$\cite{N}. The 
attractive features of the HQET are generally verified in phenomena 
where both of initial and final states include ground states of heavy 
hadron.
However, the phenomena between a ground state and an excited one, for 
example branching ratio of $B$ meson semileptonic decay into the 
excited charmed meson state $B\to D^{**}l\nu$, where $D^{**}$ means a 
higher resonance state of charmed meson, are not interpreted in 
various models of hadrons based on the 
HQET\cite{IS,CNP,SHJ,W,IS2,VO,TBS}. This discrepancy could be 
reduced to the models of hadrons as a composite system.
It is expected that the models which interpret the $B$-meson 
branching fractions of the semileptonic decays
into $D$, $D^*$  and $D^{**}$s.
At present we have no such a model.
So it is meaningful to study these processes by a model-independent 
approach.
The purpose of this short note is a test of the HQET applicability to 
an excited state
by dealing with charmed hadron excited states $D^{**}$ without model 
dependence.

In the heavy quark limit $m_Q \rightarrow \infty$, 
heavy quark spin and velocity are free from low energy QCD\cite{IW}. 
So hadron state is factorized into heavy quark $|Q\rangle$ and light 
degrees of freedom ({\it l.d.f.}) $|{ldf}\rangle$ as \begin{equation} 
|{\rm hadron}\rangle = |Q\rangle \otimes |{ldf}\rangle. 
\end{equation} These hadrons are classified by {\it l.d.f.} as 
following \begin{eqnarray}
(0^-,1^-) &=& |Q(\frac{1}{2}^+)\rangle \otimes 
|{ldf}(\frac{1}{2}^-)\rangle, \nonumber \\ (0^+,1^+) &=& 
|Q(\frac{1}{2}^+)\rangle \otimes |{ldf}(\frac{1}{2}^+)\rangle, \\
(1^+,2^+) &=& |Q(\frac{1}{2}^+)\rangle \otimes 
|{ldf}(\frac{3}{2}^+)\rangle, \nonumber \\ (1^-,2^-) &=& 
|Q(\frac{1}{2}^+)\rangle \otimes |{ldf}(\frac{3}{2}^-)\rangle, 
\nonumber
\end{eqnarray}
where we use the notation as
$$
(J_-^P,J_+^P) = |Q(\frac{1}{2}^+)\rangle \otimes |{ldf}(j^P)\rangle,
$$
with
\begin{equation}
J_\pm=j\pm \frac{1}{2} \nonumber
\end{equation}
and $J_{\pm}^P$ of left hand side denotes spin-parity of heavy 
hadrons and $j^P$ of right hand side is spin-parity of {\it l.d.f.}. 

In semileptonic decay conserving light degrees of freedom like 
$j^P=\frac{1}{2}^- \to \frac{1}{2}^-$ such as $B(0^-)$ goes to 
$D(0^-)$ or $D^*(1^-)$, there are 6 independent form factors defined 
as \begin{eqnarray}
\langle D(v')|J_{\mu}|B(v)\rangle
&=& f_+(y)(v+v')_{\mu} + f_-(y)(v-v')_{\mu}, \nonumber \\ \langle 
D^*(v')|J_{\mu}|B(v)\rangle &=&
ig(y)\epsilon_{\mu\nu\alpha\beta}
\varepsilon^{*\nu}v'^{\alpha}v^{\beta}- (y+1)f(y) \varepsilon 
^*_{\mu} \nonumber \\ & & + (\varepsilon ^*\!\cdot v) \left\{{\tilde 
a}_+(y)(v+v')_{\mu}+
{\tilde a}_-(y)(v-v')_{\mu} \right\}\ ,
\end{eqnarray}
where $v$ and $v'$ are the velocity of $B$ and $D$ or $D^*$, 
respectively. As the result of heavy quark symmetry, relations among 
these form factors are obtained and finally there exists only one 
independent form factor in these processes\cite{IW}. This form factor 
written by $\xi (y)$
is called as Isgur-Wise function (IW function). Following the HQET we
obtain only one independent form factor in each semileptonic process 
such as $j^P=\frac{1}{2}^- \to \frac{1}{2}^+$, $\frac{3}{2}^+$, 
$\frac{3}{2}^-$,$\ldots$\cite{TBS,IWe}.
In this letter we denote
these IW functions as $\xi_E^{}$, $\xi_F^{}$ and $\xi_G^{}$ for 
$j^P=\frac{1}{2}^+,\,\frac{3}{2}^+,\,\frac{3}{2}^-$, respectively. 
The relations between form factors and IW function are given in 
Ref.\cite{TBS}.

We can derive the Bjorken sum rule on IW functions\cite{IWe,B}. In 
the HQET hadron state is factorized into heavy and light degrees of 
freedom like in Eq.(1)
and the sum of transition rate between heavy hadrons $H_Q$ and 
$H'_{Q'}$ is equal to transition rate of heavy quark $Q$ into $Q'$ as 
\begin{eqnarray}
\sum_{H'_{Q'}} g^{\mu\nu} \langle H'_{Q'}|J_{\mu}|H_Q\rangle \langle 
H_{Q}|J^{\dagger}_{\nu}|H_Q'\rangle &=&
g^{\mu\nu} \langle Q'|J_{\mu}|Q\rangle
\langle Q|J^{\dagger}_{\nu}|Q'\rangle \, \sum_{ldf'}|\langle 
{ldf'}|{ldf}\rangle |^2 \nonumber \\ &=& \langle Q'|J^{\mu}|Q\rangle 
\langle Q|J^{\dagger}_{\mu}|Q'\rangle \nonumber \\ &=& -8y, 
\label{8y} \end{eqnarray}
where $y\equiv v\cdot v'$ and here we use the unitarity relation 
\begin{equation}
\sum_{ldf'}|\langle {ldf'}|{ldf}\rangle |^2 = 1. \end{equation} On 
the other hand in the case of heavy meson transition $B \to D_X$, we 
obtain
\begin{eqnarray}
\sum_{H'=D,D^*} g^{\mu\nu} \langle H'(v')|J_{\mu}|B(v)\rangle \langle 
B(v)|J^{\dagger}_{\nu}|H(v')\rangle &=& -4y(y+1)|\xi (y)|^2, 
\nonumber\\
\sum_{H'=D_0^*,D_1} g^{\mu\nu} \langle H'(v')|J_{\mu}|B(v)\rangle 
\langle B(v)|J^{\dagger}_{\nu}|H(v')\rangle &=& -4y(y-1)|\xi_E^{} 
(y)|^2, \nonumber\\ \sum_{H'=D_1,D_2^*} g^{\mu\nu} \langle 
H'(v')|J_{\mu}|B(v)\rangle \langle 
B(v)|J^{\dagger}_{\nu}|H(v')\rangle &=& 
-\frac{8}{3}y^2(y^2-1)(y+1)|\xi_F^{} (y)|^2, \nonumber\\ 
\sum_{H'=D_1^*,D_2} g^{\mu\nu} \langle H'(v')|J_{\mu}|B(v)\rangle 
\langle B(v)|J^{\dagger}_{\nu}|H(v')\rangle &=& 
-\frac{8}{3}y^2(y^2-1)(y-1)|\xi_G^{} (y)|^2, \nonumber\\ 
\sum_{H'=D_{C_2},D^*_{C_2}} g^{\mu\nu} \langle 
H'(v')|J_{\mu}|B(v)\rangle
\langle B(v)|J^{\dagger}_{\nu}|H(v')\rangle &=& -4y(y+1)|\xi_{C_2}^{} 
(y)|^2, \label{uhen} \end{eqnarray} by the straight 
calculation\cite{comm},
where index $C_2$ denotes radial excitation of $j^P$ = 
$\frac{1}{2}^-$. Here, we introduce the hypothesis of resonance 
saturation which means that it is possible to neglect the 
contribution from the continuum spectra to semileptonic decay. Under 
this assumption
the relation \begin{eqnarray}
1 &=& \frac{y+1}{2} \left(|\xi (y)|^2 + |\xi_{C_2}^{} (y)|^2 \right) 
+ \frac{y-1}{2}|\xi_E^{}(y)|^2 +
\frac{1}{3}y(y^2-1)(y+1)|\xi_F^{}(y)|^2 \nonumber \\ & & + 
\frac{1}{3}y(y^2-1)(y-1)|\xi_G^{}(y)|^2 + \cdots \label{sum} 
\end{eqnarray}
is derived by the comparison of Eqs.(\ref{8y}) and (\ref{uhen}), 
where dots means contributions from other higher resonances. This is 
the simplified Bjorken sum rule\cite{IWe,B}. 

In $B$ meson semileptonic decay, differential decay rate is given as, 
\begin{equation}
\frac{d\Gamma_X}{dy}\equiv\frac{d\Gamma}{dy}(B\to D_X l\nu) = 
\frac{G_F^2|V_{cb}|^2}{48\pi^3}
m_B^2 \sqrt{y^2-1} m_{D_X}^3 W_X(y,r^{}_X) |\xi_X^{}(y)|^2\,, 
\nonumber \end{equation}
where lepton mass is neglected and $W_X(y,r^{}_X)$ is a calculable 
functions of $y$
and $r_X^{}=\frac{m_{D_X}^{}}{m_B^{}}$\cite{TBS}. 
In the following analysis, 
We assume that the excited states, which  contribute to the 
simplified Bjorken sum rule, saturate the $B$-meson 
semileptonic decay rate(resonance saturation hypothesis) and 
that these exited states occur in a mass range being small compared 
with $m_c$. 
The latter assumption leads us to replace $m_{D_X}(X=C_2,E,F,G,\ldots)$ by 
a common mass
$m_{D^{**}}$.
Then we can sum up 
these decay rates for all $D^{**}$ and we get the equality \begin{eqnarray} 
\frac{d\Gamma_{**}}{dy} &\equiv& \frac{d\Gamma}{dy}(B\to D^{**}l\nu) 
\nonumber\\
&=& \sum_{X=C_2,E,F,G,\ldots} \frac{d\Gamma_X}{dy} \nonumber\\ &=& 
\frac{G_F^2|V_{cb}|^2}{48\pi^3} m_B^2 \sqrt{y^2-1} \sum_{X=C_2,E,F,G,\ldots}
m_{D_X}^3 
W_X(y,r^{}_X)|\xi_X^{}(y)|^2 \nonumber \\ &=& 
\frac{G_F^2|V_{cb}|^2}{24\pi^3} m_B^2 \sqrt{y^2-1} m_{D^{**}}^3 
\nonumber \\
&\times & \left[(y-1)(1+r)^2 + (y+1)(1-r)^2 + 4y(1+r^2-2ry)\right] 
\nonumber \\
&\times &
\left[\frac{y+1}{2}|\xi_{C_2}^{}(y)|^2
+ \frac{y-1}{2}|\xi_E^{}(y)|^2
+ \frac{1}{3}y(y^2-1)(y+1)|\xi_F^{}(y)|^2 \right. \nonumber\\ & & 
\left. + \frac{1}{3}y(y^2-1)(y-1)|\xi_G^{}(y)|^2 + \cdots \right].
\end{eqnarray}
By using Eq.(\ref{sum}), we obtain the $D^{**}$ contribution as 
\begin{eqnarray}
\frac{d\Gamma_{**}}{dy} &=&
\frac{G_F^2|V_{cb}|^2}{24\pi^3} m_B^2 \sqrt{y^2-1} m_{D^{**}}^3 
\nonumber \\
& & \times \left[(y-1)(1+r)^2 + (y+1)(1-r)^2 + 4y(1+r^2-2ry)\right] 
\nonumber \\
& & \times \left[1-\frac{y+1}{2}|\xi (y)|^2\right]. \end{eqnarray} 
>From this result we can estimate the $D^{**}$ contribution by using 
the following parameters\cite{PDG}
\begin{equation}
V_{cb}=0.040, \qquad
m_B=5.279 \, \mbox{GeV}, \qquad \tau_B=1.537\, ps \end{equation}
and for the mass of $D^{**}$ we use the following weighted average 
mass
\begin{equation}
m_{D^{**}}=\frac{3m_{D_1} + 5m_{D_2^*}}{8} = 2.444 \,\mbox{GeV} 
\end{equation}
with $m_{D_1}=2.420\mbox{GeV}$ and $m_{D_2^*}=2.460\mbox{GeV}$. We 
use the IW function $\xi (y)$ with following three trial functions 
\begin{tabbing}
\hspace*{13mm} \= \hspace*{35mm} \= \hspace*{30mm} \kill (I) \> 
$1-\rho^2(y-1)$ \> $\rho = 0.91{+0.19 \atop -0.21}$, \\ (II) \> 
$\exp[-\rho^2(y-1)]$ \> $\rho = 0.99{+0.27 \atop -0.25}$, \\ (III)\> 
$\left(\frac{2}{y+1}\right)^{2\rho^2}$ \> $\rho = 1.03 {+0.28 \atop 
-0.26}$,
\end{tabbing}
where $\rho$ is a free parameter to reproduce the experimental 
branching ratio
\begin{eqnarray*}
{\rm Br}(B\to D^{(*)} l\nu)
&=& {\rm Br}(B\to D l\nu)+{\rm Br}(B\to D^* l\nu) \\ &=& (1.6 \pm 
0.7)\,\% + (6.6 \pm 2.2) \,\% \\ &=& (8.2 \pm 2.3) \,\%, 
\end{eqnarray*}
and we use $m_{D^{(*)}}=\frac{m_D+3m_{D^*}}{4}=1.975 \mbox{GeV}$ to 
determine $\rho$.
\vspace*{-5mm}\\
\begin{table}[h]
\caption{Branching ratio $B\to D_Xl\nu$
$\left(|\frac{V_{cb}}{0.040}|^2\frac{\tau_{B}^{}} 
{1.537\,ps}\,\%\right)$}
\begin{tabular}{lllll}\hline
\makebox[15mm]{}
& \makebox[25mm]{$D$} & \makebox[25mm]{$D^*$} & 
\makebox[25mm]{$D^{**}$}
& \makebox[25mm]{$\sum_X D_X$} \\ \hline (I) & $1.91$ ${-0.68 \atop 
+0.76}$ & $6.24$ ${-1.53 \atop +1.63}$ & $1.58$ ${+0.89 \atop -0.86}$ 
& $9.73$ ${-1.33 \atop +1.52}$ \\ \hline
(II) & $1.95$ ${-0.67 \atop +0.72}$ & $6.22$ ${-1.61 \atop +1.61}$ & 
$1.71$ ${+1.05 \atop -0.93}$ & $9.88$ ${-1.24 \atop +1.40}$ \\ \hline
(III) & $1.96$ ${-0.67 \atop +0.72}$ & $6.21$ ${-1.62 \atop +1.62}$ & 
$1.77$ ${+1.06 \atop -0.95}$ & $9.94$ ${-1.23 \atop +1.39}$ \\ \hline
Average & $1.94$ ${-0.68 \atop +0.73}$ & $6.22$ ${-1.59 \atop +1.62}$ 
& $1.69$ ${+1.00 \atop -0.91}$ & $9.85$ ${-1.27 \atop +1.44}$ \\ 
\hline
\end{tabular}
\label{D}
\end{table}

The $D^{**}$ contribution and $D_X$ total contribution 
($D+D^*+D^{**}$) are given in Table~\ref{D}. Here we use Eq.(11) to 
obtain $D^{**}$ contribution which gives about 1.7\% branching 
fraction.
The direct measurement of $D^{**}$ contribution is $2.7\pm 
0.7\%$\cite{PDG}. This is a factor 1.6 larger at central value than 
the theoretical estimation of $D^{**}$ obtained by the simplified 
Bjorken sum rule Eq.(\ref{sum}).
However, the experimental error is still large and the estimated 
magnitude of $D^{**}$ contribution seems to be consistent with the 
present experiment.
On the other hand, inclusive semileptonic decay branching ratio is 
experimentally $10.43\pm 0.24\%$ and the unidentified semileptonic 
branching fraction \cite{PDG} is
\begin{equation}
{\rm Br}(B\to \mbox{unknown})
={\rm Br(inclusive)} - {\rm Br}(B\to D \mbox{and} D^*l\nu) = 2.2 \pm 
2.3\,\%.
\end{equation}
This also shows that the resonance saturation hypothesis and the approximate 
mass degeneracy among the excited charmed mesons($D_0^*, D_1, D_1^*, 
D_2, D_2^*, D_{c_2}, D^*_{c_2} \cdots$)  might hold 
in $B$ meson semileptonic decay.
%\vspace*{-5mm}\\
\begin{table}
\caption{Comparison of $D^{**}$ contribution 
$\left(|\frac{V_{cb}}{0.040}|^2\frac{\tau_{B}^{}}{1.537\, 
ps}\,\%\right)$}
\begin{tabular}{lrrrrrr}
\hline \makebox[20mm]{} & 
\makebox[15mm]{ISGW2\cite{IS2}} & \makebox[15mm]{SISM\cite{TBS}} & 
\makebox[15mm]{CNP\cite{CNP}} & \makebox[15mm]{VO\cite{VO}} & 
\makebox[15mm]{ours${}^{*}$}
& \makebox[15mm]{Exp.\cite{PDG}} 
\\ \hline Br($D+D^*$) & 9.03 & 7.23 & 7.24 & 9.24 & 
8.16$^\dagger$
& 8.2$\pm$ 2.3 \\ \hline
Br($D^{**}$) & 0.96 & 0.33 & 0.53 & 0.96 & 1.69 & 2.7$\pm$ 0.7 \\ 
\hline 
\end{tabular}
\begin{tabular}{l}
${}^*$In this table ours data 
are average in Table~\ref{D}.\\
${}^\dagger$This is an input value.\\
\end{tabular}
\label{D**}
\end{table}

In order to estimate the contribution from each resonance, it is 
necessary to calculate exclusive decay processes by using hadronic 
models as given in
Ref.\cite{IS,CNP,SHJ,W,IS2,VO,TBS}.
Through these studies there is a tendency that branching fractions 
into $D_2^*$ and $D_1$ are rather larger compared to the other 
excited states.
The maximum fraction among $D^{**}$s is $D_2^*$ in 
Ref.\cite{CNP,VO,TBS} with the magnitude 0.1\% $\sim$ 0.4\% and is 
$D_1$ in Ref.\cite{IS2} with 0.4\%.
To make clear which model is better,
further exclusive experiments are needed. One more feature in common 
with model-dependent analyses given in Table~\ref{D**} is that 
relatively small $D^{**}$ fraction is predicted and is inconsistent
with experimental value.

Further in order to check our model-independent approach we apply 
this method to $B_s$ semileptonic decay processes. The IW function of
$B_s \to D^{(*)}_s l\nu$ is the same one of $B\to D^{(*)} l\nu$, 
because $u$, $d$ and $s$ quarks are treated as {\it l.d.f.} in the 
HQET. In the numerical estimation the parameters are\cite{PDG} 
\begin{eqnarray*}
m_{B_s}^{} &=& 5.375\, \mbox{GeV}, \qquad \tau_{B_s}^{}=1.34\, ps, \\
m_{D_s^{(*)}}^{} &=& \frac{m_{D_s}^{} + 3 m_{D_s^*}^{}}{4} = 2.075\, 
\mbox{GeV}, \\
m_{D_s^{(**)}}^{} &=& \frac{3 m_{D_{s1}}^{} + 5 m_{D_{s2}^*}^{}}{8} = 
2.556\, \mbox{GeV}\,,
\end{eqnarray*}
where $m_{D_{s2}^*}^{}\cong 2568 \mbox{MeV}$ is a presumption from 
other charmed meson masses by using the relation $m_{D_{s2}^*}^{}$ 
$-$ $m_{D_{s1}^{}}(2535)$ = $m_{D_2^*}^{}(2456)$ $-$ 
$m_{D_1}^{}(2423)$. Following the same argument of $B\to D_Xl\nu$ we 
can give the branching ratios of $B_s$ to $D_s$, $D_s^*$ and 
$D_s^{**}$ in Table~\ref{Ds}.
The predicted branching ratio is similar to ones of $B\to D_Xl\nu$ as 
shown in Table~\ref{D}. The largest branching ratio
is reduced to a fraction to $D_s^*$ and a contribution from excited 
states $D_s^{**}$ is less than 16\% of inclusive ratio. If this is 
confirmed by experiments it will verify the validity of the analysis 
using the Bjorken sum rule.
\vspace*{-5mm}\\
\begin{table}[h]
\caption{Branching ratio $B_s\to D_{sX}l\nu$ 
$\left(|\frac{V_{cb}}{0.040}|^2\frac{\tau_{B_s}^{}}{1.34\, 
ps}\,\%\right)$} \begin{tabular}{lllll}\hline \makebox[15mm]{}
& \makebox[25mm]{$D_s$} & \makebox[25mm]{$D_s^*$} & 
\makebox[25mm]{$D_s^{**}$} & \makebox[25mm]{$\sum_X D_{sX}$} \\ 
\hline (I) & $1.79$ ${-0.59 \atop +0.64}$ & $5.78$ ${-1.31 \atop 
+1.37}$ & $1.28$ ${+0.73 \atop -0.70}$ & $8.84$ ${-1.17 \atop +1.31}$ 
\\ \hline
(II) & $1.80$ ${-0.59 \atop +0.61}$ & $5.72$ ${-1.41 \atop +1.38}$ & 
$1.40$ ${+0.88 \atop -0.76}$ & $8.92$ ${-1.12 \atop +1.23}$ \\ \hline
(III) & $1.81$ ${-0.59 \atop +0.61}$ & $5.70$ ${-1.42 \atop +1.39}$ & 
$1.45$ ${+0.89 \atop -0.78}$ & $8.96$ ${-1.12 \atop +1.22}$ \\ \hline
Average & $1.80$ ${-0.59 \atop +0.62}$ & $5.73$ ${-1.38 \atop +1.38}$ 
& $1.38$ ${+0.83 \atop -0.75}$ & $8.91$ ${-1.14 \atop +1.25}$ \\ 
\hline
\end{tabular}
\label{Ds}
\end{table}

In this letter we estimate the semileptonic branching ratios of 
$B$ to excited 
charmed mesons by using a method independent on specific hadron 
models.
The Bjorken sum rule will be checked by measuring the contribution 
>from higher resonance of charmed meson in semileptonic decay of 
$B_{u,d}$ and $B_{s}$
meson.
The prediction is given under the assumption that the semileptonic 
decay is saturated by the three body decays $B \rightarrow D_{X}l\nu$ 
and the continuum contribution is negligible and 
that the excited states have approximately equal masses.
We get Br($B\to D^{**}l\nu$) =
1.7$\pm$ 1.0\% which seems to be consistent with the experimental 
value 2.7$\pm$ 0.7\%.
We also estimate Br($B_s\to D_s,D_s^*,D_s^{**}l\nu$) to be 1.8\%, 
5.7\% and 1.4\%, respectively.
The predicted branching ratios are similar to Br($B\to 
D,D^*,D^{**}l\nu$). These evaluations are to be checked by 
experiments in near future and we expect that the model-independent 
approach will be
confirmed experimentally.

%%%%%%%%%%%%%%%%%%%%%%%%%%%%%%%%%%%%%%%%%%%%%%%%%%%%%%%%%%%%%%%%%%%%%%%% 

\vskip 1 cm
\noindent
{\bf Acknowledgments}\par
We are grateful for enjoyable discussions with T. Itoh and Y. Matsui.
One of authors(M. M) would like to thank for the financial support by 
the Grant-in-Aid for Scientific Research, Ministry of Education, 
Science and Culture, Japan(No.06640386). \vskip 1 cm
%%%%%%%%%%%%%%%%%%%%%%%%%%%%%%%%%%%%%%%%%%%%%%%%%%%%%%%%%%%%%%%%%%%%%%%%% 

%%%%%%%%%%%%%%%%%%%%%%%%%%%%%%%%%%%%%%%%%%%%%%%%%%%%%%%%%%%%%%%%%%%%%%%%% 


\begin{thebibliography}{99}
\bibitem{comm}
To obtain these equations, we use Eqs.(13a)-(13l) and the form factor 
relation given in Table II in Ref.\cite{TBS}.
\end{thebibliography}
\end{document}